\documentclass{PoS}
\usepackage{amsmath}
\usepackage{wrapfig}

\title{Large N meson propagators from twisted space-time reduced model}

\ShortTitle{Large N meson propagators from twisted space-time reduced model}

\author{Antonio Gonz\'alez-Arroyo$^{ab}$ \\
\llap{$^a$}Instituto de F\'{\i}sica Te\'orica UAM/CSIC, C/ Nicol\'as Cabrera 13-15\\
       Universidad Aut\'onoma de Madrid, E-28048--Madrid, Spain \\
\llap{$^b$}Departamento de F\'{\i}sica Te\'orica, C-15 \\
       Universidad Aut\'onoma de Madrid, E-28049--Madrid, Spain\\
E-mail: \email{antonio.gonzalez-arroyo@uam.es}}       
\author{\speaker{Masanori Okawa}$^{cd}$ \\      
\llap{$^c$}Graduate School of Science, Hiroshima University\\
Higashi-Hiroshima, Hiroshima 739-8526, Japan\\
\llap{$^d$}Core of Research for the Energetic Universe, Hiroshima University\\
Higashi-Hiroshima, Hiroshima 739-8526, Japan\\
E-mail: \email{okawa@sci.hiroshima-u.ac.jp}}

\abstract{Recently, we proposed a new method to calculate meson propagators in the large $N$ limit 
from twisted space-time reduced model. In this note, we give simulation details for obtaining
meson spectra and discuss the smearing technique which should improve the signal of meson propagators in future works.}

\FullConference{The 33rd International Symposium on Lattice Field Theory\\
		14 -18 July 2015\\
		Kobe International Conference Center, Kobe, Japan*}

\begin{document}

\section{Introduction}
In the last few years, we have performed several studies of  the twisted space-time reduced model.  
This is a one-site lattice model\cite{TEK1,TEK2}, expected to become
equivalent in the large $N$ limit to the usual lattice gauge theory at
infinite volume.  We have identified the condition on the twisted boundary condition 
which ensures this equivalence\cite{TEK3}.  We have also calculated a variety of quantities on 
both  lattice\cite{testing} and continuum\cite{string,coupling}
space-time.  We also showed that it is   straightforward to introduce adjoint fermions 
within our framework\cite{nf2prd,nf2jhep}.   

Although one of the most important observables of  lattice gauge
theories is given by the  hadron spectra, our previous studies did not
attempt to calculate it in the twisted reduced model. 
At first sight it seemed impossible  to calculate space-time extended hadron correlation 
functions in the one-site model. Furthermore, the presence of quarks
in the  fundamental representation of the group seemed in conflict with twisted boundary 
conditions.  Recently, we found a way to circumvent these difficulties 
and succeeded in calculating meson spectra in the large $N$ limit\cite{meson}.  
The purpose of this note is to present simulation details which are not covered in ref\cite{meson}, 
and propose a new smearing method which should be indispensable for a precise determination of meson spectra.          

\section{Formulation}
We consider the twisted Eguchi-Kawai model with gauge group SU($N$) with $N=L^2$, $L$ being positive 
(preferably prime) integer.  The action of the TEK model is given by
\begin{equation}
   S=-bN \sum_{\mu \ne \nu =0}^{d-1}  z_{\nu\mu} {\rm Tr}\left[ U_\mu
  U_\nu U_\mu^\dagger U_\nu^\dagger \right]      
\label{S}
 \end{equation}
Here, $U_{\mu}$ are d=4 SU($N$) link variables. The symmetric twist
tensor $z_{\mu\nu}$ is an element 
of Z($L$);
\begin{equation}
 z_{\nu\mu} = \exp \left( k {2\pi i \over L} \right), \ \ \ z_{\mu\nu}=z_{\nu\mu}^*, \ \ \ \mu>\nu 
\label{Z}
\end{equation}
$k$ and $L$ are co-prime and should obey some constraint in order to
avoid Z($N$) symmetry breaking at intermediate values of the coupling\cite{TEK3}.  
The minimum action solution is given by matrices $\Gamma_\mu$ satisfying
\begin{equation}
    \Gamma_\mu \Gamma_\nu = z_{\mu\nu}  \Gamma_\nu \Gamma_\mu      
\label{G}
 \end{equation}
whose explicit form is given, for example, in eq. (2.13) of ref.\cite{testing}. 

As has been shown in ref.\cite{meson}, the meson propagator in channel $\gamma_A$ and $\gamma_B$ at time distance $n_0$ is given by the following formula  
\begin{equation}
C_{AB}(n_0)={1 \over \ell_0 N^{3/2}} \sum_{q_0}{\rm e}^{-i q_0 n_0} {\rm Tr} \left[\gamma_A D^{-1}(0)\gamma_B D^{-1}(q_0)\right]
\label{meson_prop}
\end{equation}
Here $D(q_0)$ is the Wilson-Dirac matrix acting on color ($U_\mu$), spatial ($\Gamma_\mu$) and Dirac ($\gamma_\mu$) spaces and its explicit form is
\begin{equation}
 D(q_0)=1-\kappa\sum_{\mu=0}^{d-1}\left[(1-\gamma_\mu){\tilde U_\mu} \Gamma_\mu^{*} +
(1+\gamma_\mu){\tilde U_\mu^\dagger} \Gamma_\mu^{t} \right]
\label{quark_action}
\end{equation}
with
\begin{equation}
{\tilde U_{\mu=0}}={\rm e}^{i q_0}U_{\mu=0},\ \ \ \ \ {\tilde U_{\mu=1,2,3}}=U_{\mu=1,2,3}
\end{equation}
and
\begin{equation}
q_0={2\pi m \over \ell_0 L},\ \ \ \ 0\le m \le \ell_0 L-1
\end{equation}

Quark fields are supposed to live in a finite box $\ell_0 L \times L^3$ with positive integer 
$\ell_0$, whereas gauge fields live in a $L \times L^3$.   A value of
$\ell_0>1$ is very convenient since in formula (\ref{meson_prop}) we represent the
correlation function  of ultralocal meson operators in spatially zero momentum state.
This correlator receives important contributions of higher excitation
and one must go to relatively long time separation to extract a good
signal for the lowest mass state.
As will be discussed in sect. 4, this problem can  be circumvented by
using spatially extended smeared meson operators.

We use $Z(4)$ random source method to compute Tr in (\ref{meson_prop}).  
Let $z(\alpha,\beta,i)$ be the source vector having color ($U_\mu$) index $1\le \alpha \le N$, spatial ($\Gamma_\mu$) index 
$1\le \beta \le N$ and Dirac ($\gamma_\mu$) index $1\le i \le 4$.
Then 
$z(\alpha,\beta,i)={1 \over \sqrt{2}}(\pm1 \pm i)$.  After averaging over random source, we have 
$\left< z^{*}(\alpha',\beta',i')z(\alpha,\beta,i)\right>=\delta_{\alpha'\alpha}\delta_{\beta'\beta}\delta_{i'i}$.  Now we define Hermitian operator $Q(q_0)=D(q_0)\gamma_5$ and solve the following $\ell_0 L + 1$ equations.
\begin{eqnarray}
&&Q(0)x_A=\gamma_5 \gamma_A^{\dagger} z \\
&&Q(q_0) \gamma_5 \gamma_B^{-1}y_B(q_0) = z,\ \ \ \ q_0={2\pi m \over \ell_0 L},\ \ \ \ 0\le m \le \ell_0 L-1
\end{eqnarray}
Taking the inner product of $x_A^{\dagger}$ and $y_B(q_0)$ and averaging over random source, we have
\begin{eqnarray}
\nonumber
\left<x_A^{\dagger} y_B(q_0) \right>&=&\left<z^{\dagger}\gamma_A \gamma_5 Q^{-1}(0) \gamma_B \gamma_5 Q^{-1}(q_0)z\right> \\
&=&{\rm Tr} \left[\gamma_A D^{-1}(0)\gamma_B D^{-1}(q_0)\right]
\label{average}
\end{eqnarray}
We use the CG inversion, so $Q^{-1}$ actually means $Q^{-1}=QQ^{-2}$.  It should be noted that once we have calculated $y_5(q_0)$ for all $q_0$, $y_B(q_0)$ can be obtained without the CG inversion as
\begin{equation}
y_B(q_0)=\gamma_B \gamma_5 y_5(q_0)
\end{equation}

\section{Simulation}
The $d=4$ gauge link variables $U_\mu$ are generated by a 
recently proposed over-relaxation Monte Carlo method\cite{OR}. 
This method produces independent gauge configuration approximately 
twice faster than the conventional heat bath method\cite{FH}.  We choose $N=289=17^2$.  Therefore 
the effective lattice size of the gauge field is $17^4$.  
We take $k=5$ to ensure that our theory
does not suffer from Z($N$) symmetry breaking and use 5 random sources to approximate (\ref{average}).  We study two values of gauge coupling $b=0.36$ and $0.37$, and for each $b$, 800 gauge configurations are stored.  Each configuration are separated by 1000 MC sweeps.  Errors are estimated by jackknife method.

In fig. 1, we show ${1 \over N}{\rm Tr} \left[\gamma_5 D^{-1}(0)\gamma_5 D^{-1}(q_0)\right]$ with  
$q_0={2\pi m \over \ell_0 L}$ for $\ell_0=2$, at $b$=0.36 and $\kappa$=0.155.   The actual error bars are too small to see in this scale, so they are artificially enhanced 100 times.  We notice error bars are larger for smaller $q_0$ (mod $2\pi$).

\begin{figure}[t]
\begin{center}
\includegraphics[width=80mm]{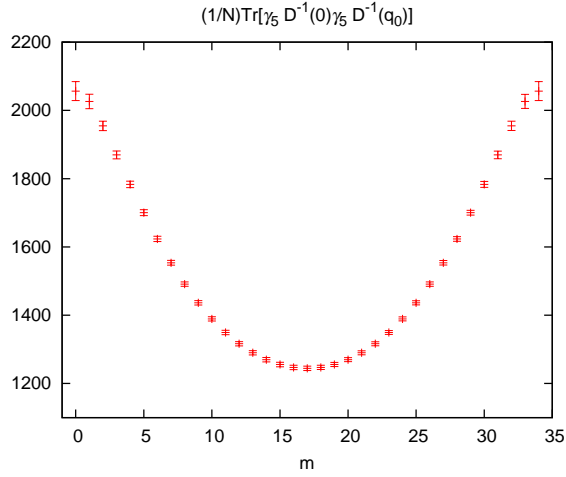}
\end{center}
\caption{${1 \over N}{\rm Tr} \left[\gamma_5 D^{-1}(0)\gamma_5 D^{-1}(q_0={2\pi m \over \ell_0 L})\right]$ as a function of $m$ at $b$=0.36 and $\kappa$=0.155.  The error bars of $y$-axis are 100 times enhanced artificially.}
\label{fig_cp34}
\end{figure}

Meson correlation functions are extracted from (\ref{meson_prop}).  By making the sum over 
$q_0={2\pi m \over \ell_0 L}$, $0 \le m \le \ell_0 L-1$ for $0 \le n_0 \le \ell_0 L-1$, we obtain the correlation function for the quarks living in a $\ell_0 L \times L^3$ finite box.
 
In fig. 2, we show the pion correlation function $C_{55}(n_0)$ both for $\ell_0 = 1$ and $2$.
For $\ell_0$ =1, signals are good for entire range of $0 \le n_0 \le L=17$.  However, contaminations from higher excited states are so large that we can not reliably extract the information of the lowest state.  On the other hand, for $\ell_0$ =2, although the data points are  
quite smooth as a function of $n_0$, the error bars become large for larger separation.   This is in sharp contrast to the convention calculation of pion correlator, where we know that the error of the pion correrator decreases proportional to the value of the correlator itself.
The reason is simple; we extract the correlation function from (\ref{meson_prop}).  As is pointed out earler, largest error comes from  ${\rm Tr} \left[\gamma_5 D^{-1}(0)\gamma_5 D^{-1}(0)\right]$, whose contribution to $C_{AB}(n_0)$ is independent on $n_0$.   

\begin{figure}[t]
\begin{center}
\includegraphics[width=80mm]{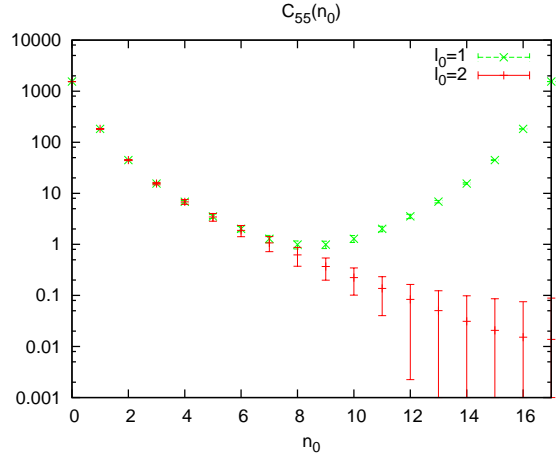}
\end{center}
\caption{Pion correlation function $C_{55}(n_0)$ as functions of $n_0$ at $b$=0.36 and $\kappa$=0.155.  Green symbols represent the results with $\ell_0=1$ whereas red symbols are for $\ell_0=2$.}
\label{fig_C_55}
\end{figure}

We, therefore, expect that constant error contributions are canceled out in effective mass $m_{\rm eff}(n_0)=\log \left[ C(n_0-1)/C(n_0) \right]$ and get good signals even for $\ell_0=2$.
Results are shown in fig. 3 for both pion (black) and rho meson(red) at $b$=0.36 and $\kappa$=0.155.
We clearly see good signals up to $n_0 \sim 13$, although contributions of higher excitation states
are not negligible for small $n_0$ due to the use of local meson operators.   We then make a three mass fit of the form
\begin{equation}
m_{\rm eff}(n_0) = \log\left[ G(n_0 -1)/G(n_0) \right]
\label{m_eff}
\end{equation}
\begin{equation}
G(n_0) = {\rm e}^{-m_1 n_0} + a_2{\rm e}^{-m_2 n_0} + a_3{\rm e}^{-m_3 n_0} 
\end{equation}
with the fitting range $[3:13]$ for pion and $[2:13]$ for rho meson.  
The fits are reasonable with $\chi^2/$ndf 0.76 for pion and 0.63 for rho meson.
We have repeated the above procedure for other $\kappa$ values with $b$=0.36 and 0.37, and 
these results are used to make discussion in ref.\cite{meson}.

\begin{figure}[t]
\begin{center}
\includegraphics[width=80mm]{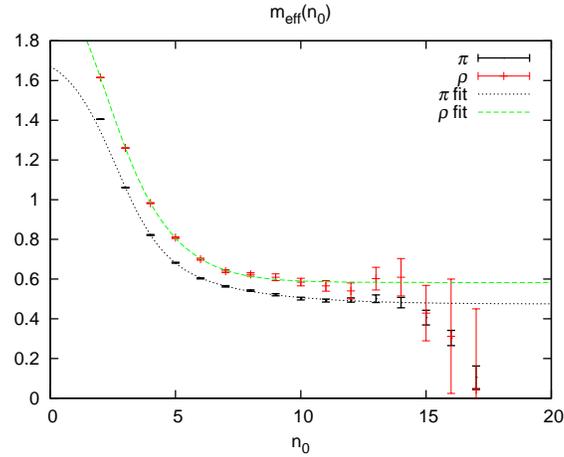}
\end{center}
\caption{Effective mass $m_{\rm eff}(n_0)=\log \left[ C(n_0-1)/C(n_0) \right]$ as functions 
of $n_0$ for pion (black) and rho meson(red).  Fits are made with fitting function (3.1). }
\label{fig_m_eff}
\end{figure}

\section{Smearing}

From the above discussion, it is obvious that we have to reduce the effects of higher excitation states to get more reliable results.  Usually, this is achieved by constructing spatially extended operators having the same quantum numbers and make a variational analysis.  At first sight,
it does not seem to be clear how to make spatially extended operators in our one-site model.  Nevertheless, 
it is a rather straightforward generalization of our formulation to construct smeared operators.

Our proposal is summarized as follow.  For the meson operator in channel $A$, 
we replace $\gamma_A$ by the following operator having the same quantum number
\begin{equation}
\gamma_A \rightarrow D_s^l \gamma_A,\ \ \ D_s \equiv {1 \over 1+6c}\left[1+c \sum_{i=1}^{d-1}\left( {\bar U_i} \Gamma_i^{*} +
{\bar U_i^\dagger} \Gamma_i^{t} \right) \right]
\label{smearing}
\end{equation}
$l$ is the smearing level and ${\bar U_i}$ is the ape-smeared spatial link variables 
after making the following transformation several times iteratively
\begin{equation}
U'_i = {\rm Proj_{SU(N)}} \left[ (1-f)U_i + {f \over 4} \sum_{j \ne i} ( U_j U_i U_j^{\dagger} + U_j^{\dagger} U_i U_j) \right]
\label{ape}
\end{equation}
$c$ and $f$ are smearing parameters. (\ref{smearing}) is just the one-site version of the smearing adopted in \cite{Bali}.  The smeared meson propagators are given by 
\begin{equation}
C_{AB}^{ll'}(n_0)={1 \over \ell_0 N^{3/2}}\sum_{q_0}{\rm e}^{-i q_0 n_0} {\rm Tr} \left[D_s^l \gamma_A D^{-1}(0)D_s^{l'} \gamma_B D^{-1}(q_0)\right],
\label{smeared_meson_prop}
\end{equation} 
which can be easily calculated from
\begin{eqnarray}
x_A^l&=&Q^{-1}(0)D_s^l \gamma_5 \gamma_A^\dagger z \\
y_B^{l'}(q_0)&=&D_s^{l'} y_B(q_0)
\end{eqnarray}

\begin{figure}
\begin{center}
\includegraphics[width=80mm]{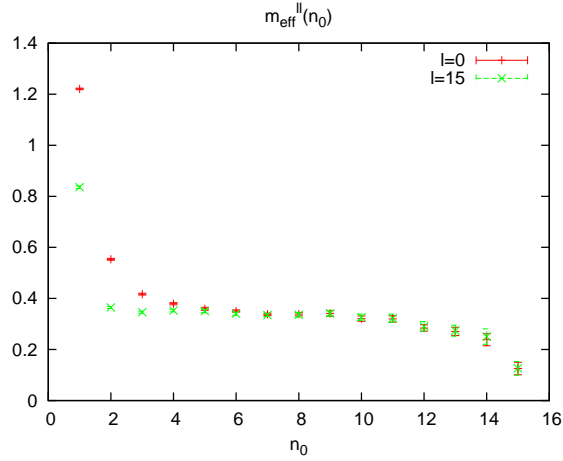}
\end{center}
\caption{Effective mass $m_{\rm eff}^{ll}(n_0)=\log \left[ C_{55}^{ll}(n_0-1)/C_{55}^{ll}(n_0) \right]$ for pion as functions of $n_0$ in $d=2$ TEK model.  Red symbols are for unsmeared operator and green symbols are 15 times smeared operator.}
\label{fig_m_eff_2d}
\end{figure}

To check the validity of the smearing, we have applied it to the $d=2$ TEK model.  In two dimension,
SU($N$) TEK model is related to the usual lattice theory on a $N \times N$ finite box.  Since there is no extra spatial dimension to make ape-smearing, $D_s$ takes very simple form
\begin{equation}
D_s = {1 \over 1+2c} \left[ 1+c\left( U_1 \Gamma_1^{*} +U_1^\dagger \Gamma_1^{t} \right) \right ]
\end{equation}   
Simulations are made with $N$=31, $k$=7, $b$=2.0 and $\kappa$=0.25.

In fig. 4, we show the effective mass of pion both for unsmeared local operator ($l=0$, red symbol)  
and that of 15 times smeared spatially extended operator ($l=15$, $c=0.4$, green symbol) with $\ell_0=1$.
The effect of the smearing is significant.  We are now making systematic variational analysis for both 2 and 4 dimensional TEK model\cite{withmarga}.  We hope to present new results in coming lattice conference.

\acknowledgments{
We acknowledge financial support from the MCINN
grants FPA2012-31686 and FPA2012-31880,
and the Spanish MINECO's ``Centro
de Excelencia Severo Ochoa'' Programme under grant
SEV-2012-0249. M. O. is supported by the Japanese MEXT grant No
26400249 and the MEXT program for promoting the enhancement of research
universities.
Calculations have been done on Hitachi SR16000 supercomputer
both at High Energy Accelerator Research Organization(KEK) and YITP in
Kyoto University. Work at KEK is supported by the Large Scale Simulation
Program No.15/16-04.

\end{document}